\documentclass{article}
\usepackage{spconf,amsmath,amssymb,graphicx}
\usepackage{etoolbox}
\apptocmd{\thebibliography}{\setlength{\itemsep}{0pt}\setlength{\parsep}{0pt}\setlength{\parskip}{0pt}}{}%
  {\PackageError{paper}{Could not set bibliography item spacing}{}}
\apptocmd{\small}{%
  \setlength{\abovedisplayskip}{2.5pt}%
  \setlength{\belowdisplayskip}{2.5pt}%
  \setlength{\abovedisplayshortskip}{2.5pt}%
  \setlength{\belowdisplayshortskip}{2.5pt}}{}%
  {\PackageError{paper}{Could not set equation spacing}{}}
\makeatletter
\patchcmd{\@maketitle}{\vskip 1.5em}{\vskip 1em}{}{\PackageError{paper}{title spacing failed}{}}
\patchcmd{\@maketitle}{\vskip 1.5em}{\vskip 1em}{}{\PackageError{paper}{title spacing failed}{}}

\renewcommand{\section}{\@startsection{section}{1}{\z@}{-6pt}{1.5pt}{\normalfont\large\bfseries}}
\patchcmd{\@maketitle}{\large}{\fontsize{12}{14}\selectfont}{}%
  {\PackageError{paper}{Could not set title font size}{}}
\patchcmd{\@maketitle}{\large}{\small}{}%
  {\PackageError{paper}{Could not set affiliation font size}{}}
\patchcmd{\@maketitle}{\@name}{\fontsize{12}{14}\selectfont\@name}{}%
  {\PackageError{paper}{Could not set author font size}{}}
\renewcommand{\subsection}{\@startsection{subsection}{2}{\z@}%
  {-3pt}{1.5pt}{\normalfont\large\bfseries}}
\patchcmd{\@makecaption}{\vskip 10pt}{\vskip 1pt}{}%
  {\PackageError{paper}{Could not set figure-to-caption spacing}{}}
\makeatother
\usepackage{hyperref}
\hypersetup{hidelinks}
\newcommand{\bx}{\mathbf{x}}
\newcommand{\bv}{\mathbf{v}}
\newcommand{\bV}{\mathbf{V}}
\newcommand{\bd}{\mathbf{d}}
\newcommand{\bg}{\overline{\mathbf{g}}}
\newcommand{\bM}{\mathbf{M}}
\newcommand{\bW}{\mathbf{W}}
\newcommand{\bL}{\mathbf{L}}
\newcommand{\bA}{\mathbf{A}}
\newcommand{\bh}{\mathbf{h}}
\newcommand{\by}{\mathbf{y}}
\title{Covariance Eigenspace Provides Latent ATTRIBUTION OF LONGITUDINAL EFFECTS IN Brain Age Gap}
\name{Jason Scheffel and Saurabh Sihag, $^*$\thanks{$^{\ast}$ for the Alzheimer's Disease Neuroimaging Initiative (see Acknowledgement section)}}
\address{University at Albany, State University of New York, Albany, NY, USA}
\begin{document}
\maketitle
% All remaining text, headings, floats and references use 9 points.
% Keep the font at 9 pt; reduce baseline-to-baseline spacing from 11 pt
% to 9.034 TeX pt (8 lines/inch).
\let\normalsize\small
\normalsize
%\linespread{0.82125}\selectfont
% Leave spare space at column bottoms instead of stretching internal gaps.
\raggedbottom

\begin{abstract}
Brain age gap is a promising machine learning (ML)-driven bio\-marker of accelerated biological aging and is derived from neuroimaging data. Recent works on coVariance neural networks (VNN) for brain age gap prediction have provided an explainable and anatomically verifiable framework for brain age gap prediction. Notably, VNN outcomes are derived by filtering inputs along covariance eigenvectors, providing a mechanism for explaining their decisions in terms of latent (orthogonal) vectors. In this paper, we investigate the longitudinal brain age gap effects in a Mild Cognitive Impairment (MCI) cohort using a VNN-driven brain age gap prediction framework. Specifically, using Integrated Gradients (IG) as the attribution mechanism, our results demonstrate that the eigenspectrum of the anatomical covariance matrix explains the different longitudinal effects observed in brain age gap in amyloid positive vs amyloid negative subcohorts more prominently than individual brain regions ($83.2\%$ with the leading eigenvector versus a maximal value of $7.7\%$ for any brain region). Therefore, the covariance-based latent explanation identifies a key coordinated input pattern that carries most of the observed longitudinal effect in brain age gap.

%predictions to eigenvector and regional coordinates; one fitted mixed model combines these scan-level contributions into its group-slope difference, with joint uncertainty. We show why predicted-age IG reconstructs the adjusted slope difference in brain age gap, defined as age-bias-corrected predicted age minus chronological age. In 464 Alzheimer's Disease Neuroimaging Initiative participants with mild cognitive impairment, we explain the amyloid-positive-minus-negative slope difference. No individual region contributes more than 7.7\% of the signed 0.375 years/year difference, whereas the leading covariance eigenvector contributes 83.2\%. Removing that direction from the fixed VNN's inputs lowers the difference to 0.062 years/year. The latent explanation therefore identifies a coordinated input pattern carrying most of the model's longitudinal effect.
\end{abstract}

\section{Introduction}
Brain age prediction models apply machine learning (ML) techniques to neuroimaging data in order to estimate how much a person's brain appears to have aged beyond their actual chronological age~\cite{franke2019ten, Sihag2025Perspective}. This discrepancy is commonly called the brain age gap. Such models have attracted growing interest in digital health and precision medicine because they offer a concise yet meaningful summary measure of brain health that could be useful in clinical settings~\cite{baecker2021machine}. The underlying premise is that these models detect atypical imaging patterns linked to biological factors (such as neurodegenerative conditions, obesity, etc.) and translate them into a biomarker that reflects the linked accelerated biological aging in a given individual. Several recent studies support this idea, demonstrating that the brain age gap (also known as the brain age delta) correlates with, and can predict, the severity of underlying biological phenomenon triggering accelerated aging~\cite{Jirsaraie2023}. Longitudinal studies are essential for determining whether the brain age gap tracks dynamic changes in brain health or simply reflects stable individual differences~\cite{franke2012longitudinal}. For instance, previous studies have reported that elevated baseline brain age gaps can be predictive of individuals with mild cognitive impairment (MCI) who later converted to Alzheimer's disease (AD)~\cite{gaser2013brainage}. In multiple sclerosis, the brain age gap was reported to be widening over time and predictive of disability progression~\cite{hogestol2019cross}. In population cohorts, larger gaps have been associated with higher mortality in older adults~\cite{cole2018brain} and with faster biological aging and cognitive decline in midlife~\cite{elliott2021brain}. Together, the aforementioned studies and numerous others have established the prognostic potential of the brain age gap. 

Despite their promise, brain age gap prediction models often function as ``black boxes," offering little insight into which brain regions or imaging features drive an individual's brain age gap. The problem is compounded in longitudinal analyses. Without interpretable models, changes in the brain age gap over time cannot be reliably traced to specific meaningful patterns, thus limiting both mechanistic insight and clinical trust in the brain age gap as a marker of disease progression. {\emph Emerging evidence suggests that latent representations may be central to both interpreting the brain age gap and capturing its longitudinal dynamics.} Covariance neural networks (VNNs) show that anatomical interpretability of an elevated brain age gap depends on the model's ability to exploit specific eigenvectors of the anatomical covariance matrix, which links the gap to identifiable brain regions~\cite{Sihag2023NeurIPS, Sihag2025Perspective}. Separately, longitudinal studies indicate that modeling change directly in a learned latent space outperforms differencing cross-sectional brain age estimates~\cite{ouyang2023lsor}. However, these two lines of work remain disconnected. Covariance-driven interpretability has been established only cross-sectionally, while longitudinal latent-space models lack an anatomically grounded explanation of what they encode. Whether longitudinal changes in the brain age gap are carried by the latent spaces within brain structure remains an open question.

%In contrast, {\bf this paper investigates whether longitudinal changes in the brain age gap can be explained by specific covariance eigenvectors}.

Recent work using coVariance neural networks (VNNs), graph neural networks that operate on the sample covariance matrix of cortical features~\cite{Sihag2022}, have provided key mathematical and operational principles to address the interpretability limitations of brain age gap~\cite{Sihag2025Perspective}. Specifically, because the information processing operations within VNNs align with the principal components of the covariance matrix~\cite{Sihagspm}, the brain age gap they predict can be traced to specific brain regions and principal components of the anatomical covariance matrix~\cite{Sihag2023NeurIPS, Sihag2025ISBI}. For instance, an elevated brain age gap in AD was driven by contributions from anatomical regions consistent with known patterns of neurodegeneration~\cite{Sihag2023NeurIPS}. In this paper, we ask whether a longitudinal effect in the brain age gap can be traced to the latent structure that VNNs use to make predictions. VNN filters are matrix polynomials of the covariance matrix and therefore act by rescaling projections of the input onto the covariance eigenvectors~\cite{Sihag2022}. These eigenvectors define a latent coordinate system shared by all filters in the network, in which each direction describes a coordinated pattern of cortical thickness across brain regions~\cite{Sihag2025ISBI, Sihag2023NeurIPS}. Existing VNN-based studies of brain age gap are primarily cross-sectional, where the brain age gap is explained by comparing the VNN final-layer regional outputs with covariance eigenvectors. Limited prior work on longitudinal analysis of brain age gap inferred by VNNs modeled brain age gap trajectories with linear mixed models~\cite{Scheffel2026EMBC}, but did not attribute those trajectories to specific input patterns.

In this paper, we focus on investigating the latent attribution of an observed clinically motivated longitudinal effect in brain age gap derived from a VNN model from structural magnetic resonance imaging (MRI) scans: the difference in annual brain age gap (denoted by $\Delta$-Age) change between amyloid-positive and amyloid-negative individuals with mild cognitive impairment (MCI) in the Alzheimer's Disease Neuroimaging Initiative (ADNI)~\cite{Jack2008}. This effect is estimated as an adjusted group-by-time coefficient in a linear mixed model that accounts for participant covariates and dependence between repeated visits. Explaining such an effect differs from explaining individual predictions. A latent component that shifts every prediction by the same amount contributes nothing to a slope difference, and any valid explanation must preserve the model's covariate adjustment and weighting of repeated visits. Attribution of statistics computed from predictions has precedents, including differences between group means~\cite{Bowen2020}, prediction–outcome correlations~\cite{Lelievre2024}, model pipelines~\cite{Chen2022} and two-sample tests~\cite{Javanbakhat2026}. To our knowledge, no existing approach decomposes an adjusted longitudinal coefficient whose observation weights come from a fitted mixed model.

Our approach rests on two observations. First, Integrated Gradients (IG)~\cite{Sundararajan2017} divides the difference between a scan's prediction and a reference prediction among input coordinates, and it can be computed in covariance coordinates using the subspace attribution construction~\cite{Chormai2024} as well as in regional coordinates. Second, once the covariance parameters of the mixed model are fixed, the estimated group-slope difference is a generalized least-squares estimate and is therefore linear in the $\Delta$-Age outcomes. Per-scan IG contributions can then be passed through the fitted model's observation weights, so that the contributions of individual latent directions sum to the reported longitudinal effect. Applied to 464 ADNI participants with MCI, this analysis shows that the leading covariance eigenvector accounts for 83.2$\%$ of the adjusted slope difference, whereas no single brain region accounts for more than 7.7$\%$. Interventions that retain or remove this direction from the VNN inputs reproduce the attribution closely, and we report that the result is robust to participant resampling, training initialization and preprocessing.

\textbf{Contributions:} Our contributions are as follows. \textbf{(i) From latent directions to longitudinal contributions.} We combine full-network IG in covariance coordinates with one fitted mixed model to decompose the adjusted group-slope difference. The contributions preserve covariate adjustment and repeated-visit weighting and sum to the reported effect up to integration error; regional IG provides a complementary decomposition. \textbf{(ii) Connecting predicted age to $\Delta$-Age.} We show that the reference prediction, chronological age and linear age-bias correction contribute zero to this coefficient under the specified design and fitted observation weights. This establishes why predicted-age IG can explain the adjusted $\Delta$-Age effect. \textbf{(iii) Uncertainty for latent contributions.} We estimate joint uncertainty for individual directions' contributions and their sums, accounting for repeated scans and dependence between components, conditional on the fitted prediction pipeline and observation weights. \textbf{(iv) Evaluating the latent explanation.} In ADNI, we compare how the same effect is distributed across latent and regional coordinates, examine the dominant direction through fixed-VNN input retention and removal, and assess sensitivity to participant sampling, training initialization and preprocessing.

\section{Explaining a longitudinal VNN effect}
\subsection{Covariance coordinates of the VNN}
Let $\bx\in\mathbb R^p$ denote one scan's cortical thickness in $p$ regions, each region standardized by its training-set mean and standard deviation. In our data (Section~\ref{sec:data}), each scan has $p=68$ regions (Desikan-Killiany atlas). With training scans as rows of $\mathbf X$, use the trace-normalized covariance
\begin{equation}
 \mathbf C=\frac{\mathbf X^{\mathsf T}\mathbf X}{\operatorname{tr}(\mathbf X^{\mathsf T}\mathbf X)}
 =\bV\boldsymbol\Lambda\bV^{\mathsf T}.
 \label{eq:cov}
\end{equation}
The columns $\bv_k$ of $\bV$ are orthonormal eigenvectors, ordered by decreasing eigenvalue $\lambda_k$ in $\boldsymbol\Lambda$. Trace normalization rescales eigenvalues without changing these directions. A covariance filter with $K$ learned coefficients $w_m$ is the matrix polynomial $\mathbf F(\mathbf C)$ in Eq.~\eqref{eq:filter}, whose second line shows that the filter acts on $\bx$ by rescaling each eigenvector projection by a factor determined by the eigenvalue~\cite{Sihag2022}:
\begin{equation}
 \begin{split}
 \mathbf F(\mathbf C)&=\sum_{m=0}^{K-1}w_m\mathbf C^m,\\
 \bv_k^{\mathsf T}\mathbf F(\mathbf C)\bx
 &=\left(\sum_{m=0}^{K-1}w_m\lambda_k^m\right)\bv_k^{\mathsf T}\bx.
 \end{split}
 \label{eq:filter}
\end{equation}
Every filter shares these directions, with learned gains determined by its coefficients $w_m$. The projections $\bV^{\mathsf T}\bx$ therefore give latent coordinates tied to the VNN's filtering structure. In the context of brain age gap prediction, VNN is a map $f:\mathbb R^p\to\mathbb R$ from a scan $\bx$ to its predicted age, built from stacked layers and pre-trained on healthy population (CN). Each VNN layer applies a filter of the form in Eq.~\eqref{eq:filter} to each input--output channel pair, sums the filtered channels with a learned bias, and applies a pointwise non-linearity (leaky ReLU in our case, Section~\ref{sec:data}). The predicted age $f(\bx)$ is the average of the final layer's outputs over channels and regions. The activations act on regional values and can mix eigenvector directions, so we compute IG through the full network, including its activations and final averaging.

\subsection{IG in regional and covariance coordinates}
IG explains the difference between a scan's prediction and the prediction at a fixed reference input, splitting that difference across input coordinates~\cite{Sundararajan2017}. Fix the VNN and let $\bx_0$ be the mean training input from the CN cohort. Standardization makes $\bx_0=0$, but $f(\bx_0)$ need not be zero. Write $\bd=\bx-\bx_0$. Along the straight path from reference to scan, define the average gradient and regional IG:
\begin{equation}
 \begin{split}
 \bg&=\int_0^1\nabla f(\bx_0+\alpha\bd)\,d\alpha,\\
 I_r^{\rm reg}&=d_r\overline g_r,\quad r=1,\ldots,p.
 \end{split}
 \label{eq:ig}
\end{equation}
Each contribution multiplies displacement by average network sensitivity. It is in predicted-age years; positive values raise predicted age relative to the reference. A region's attribution therefore depends on the network's average sensitivity to that region, not only on how far its thickness lies from the reference.

For eigenvector IG, express the displacement as $\mathbf z=\bV^{\mathsf T}\bd$ and $\widetilde f(\mathbf z)=f(\bx_0+\bV\mathbf z)$. Each latent coordinate $z_k=\bv_k^{\mathsf T}\bd$ measures displacement from the reference along direction $\bv_k$. The chain rule gives $\nabla\widetilde f(\mathbf z)=\bV^{\mathsf T}\nabla f(\bx_0+\bV\mathbf z)$. IG along the same input path $\alpha\mathbf z$ gives
\begin{equation}
 \begin{split}
 I_k^{\rm eig}
 &=z_k\int_0^1[\nabla\widetilde f(\alpha\mathbf z)]_k\,d\alpha
 =(\bv_k^{\mathsf T}\bd)(\bv_k^{\mathsf T}\bg),\\
 \sum_{k=1}^p I_k^{\rm eig}
 &=\bd^{\mathsf T}\bg=\sum_{r=1}^p I_r^{\rm reg}=f(\bx)-f(\bx_0).
 \end{split}
 \label{eq:spectral}
\end{equation}
Completeness, the property that the attributions sum to $f(\bx)-f(\bx_0)$ as in the second line of Eq.~\eqref{eq:spectral}, follows by integrating the path derivative, which exists almost everywhere for this continuous piecewise-linear VNN~\cite{Sundararajan2017}. We rotate displacement and gradient separately, not the regional IG vector, so eigenvector IG is not $\bV^{\mathsf T}$ applied to the regional IG vector. Reversing an eigenvector's sign reverses both factors, leaving its attribution unchanged.

In our experiments, the leading eigenvalue of the anatomical covariance matrix is well separated, but many smaller eigenvalues are close to their neighbours, so covariance re-estimation can rotate those directions~\cite{DavisKahan1970}. Subspace IG~\cite{Chormai2024} groups a set $U$ of directions as $I_U=\bd^{\mathsf T}\mathbf P_U\bg$, where $\mathbf P_U=\sum_{k\in U}\bv_k\bv_k^{\mathsf T}$; the sum is independent of the basis within this span. We use this for the aggregate of eigenvectors 2 to 68 in Section~\ref{sec:contrib}.

\subsection{From predicted age to $\Delta$-Age}
IG above explains predicted age, whereas the longitudinal model analyzes $\Delta$-Age. We first connect these quantities at each scan, then show why their adjusted group-slope coefficients coincide. Let $a$ be the chronological age at a scan. In CN participants, we regress predicted age minus chronological age, $f(\bx)-a$, on $a$~\cite{beheshti2019bias}, using out-of-fold predictions so that each participant's prediction comes from a model trained without that participant. Let $\widehat\gamma_0,\widehat\gamma_1$ be the fitted intercept and slope. Brain age is the corrected prediction $f(\bx)-\widehat\gamma_0-\widehat\gamma_1a$~\cite{Sihag2023NeurIPS}, and $\Delta$-Age for a scan is
\begin{equation}
 y=f(\bx)-b(a),\qquad b(a)=\widehat\gamma_0+(1+\widehat\gamma_1)a.
 \label{eq:gap}
\end{equation}
Here $b(a)$ is the predicted age expected for a CN participant of age $a$, and $y$ is the excess over it. Along a scan's IG path, age and correction stay fixed, so $\nabla_{\bx}y=\nabla_{\bx}f$. Subtracting the same $b(a)$ from prediction and reference preserves their difference: predicted-age IG also decomposes $y-y_0$, where $y_0=f(\bx_0)-b(a)$.

\begin{figure*}[t]
 \centering
 \includegraphics[width=0.8\textwidth]{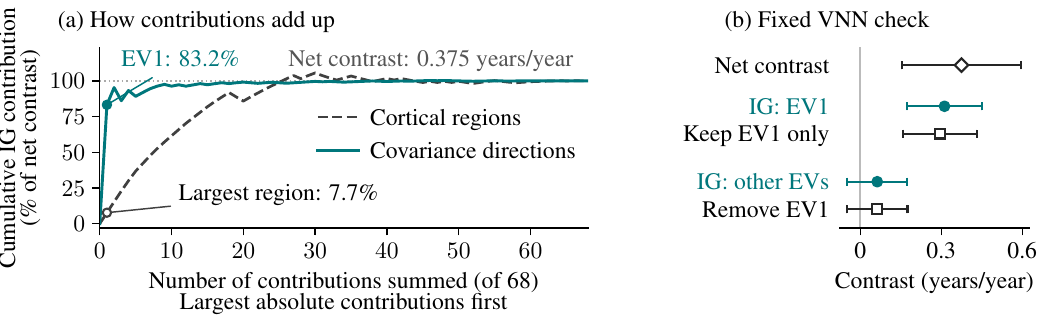}
 \caption{IG decompositions of the adjusted group-slope contrast. (a) Within each coordinate system, all 68 components are ordered by decreasing absolute contribution and their signed estimates are summed; negative contributions lower the curves. Percentages use the signed net contrast of 0.375 years/year. (b) Total contrast (diamond), IG contributions (circles) and fixed-VNN interventions (squares), with participant-clustered 95\% CR1 confidence intervals using the original observation weights. ``Other EVs'' denotes the EV2--68 sum.}
 \label{fig:results}
\end{figure*}

\subsection{One mixed-model fit for all contributions}
The linear mixed model gives each participant their own baseline $\Delta$-Age and annual slope, as random effects around group-level trends. For participant $i$ at visit $j$, let $y_{ij}$ be $\Delta$-Age, $t_{ij}$ years from baseline, and $D_i=1$ for baseline amyloid positivity, $0$ otherwise. Let $\mathbf u_i$ contain baseline age, sex, and education. We adjust both baseline $\Delta$-Age and slopes for these covariates~\cite{Laird1982}:
\begin{equation}
 \begin{split}
 y_{ij}={}&\beta_0+\beta_t t_{ij}+\beta_D D_i+\theta D_i t_{ij}\\
 &+\boldsymbol\beta_u^{\mathsf T}\mathbf u_i
 +\boldsymbol\beta_{ut}^{\mathsf T}\mathbf u_i t_{ij}
 +b_{0i}+b_{1i}t_{ij}+\epsilon_{ij}.
 \end{split}
 \label{eq:lmm}
\end{equation}
The target $\theta$ is the adjusted positive-minus-negative slope difference, in years/year. Participant effects $(b_{0i},b_{1i})$ are independent across participants, with mean zero and covariance $\mathbf G$; residuals $\epsilon_{ij}$ are independent of these effects and each other, with mean zero and variance $\sigma^2$. We fit this Gaussian model by restricted maximum likelihood (REML), then fix its longitudinal covariance parameters when analyzing components.

Stack the $\Delta$-Age values of the $n$ scans from $N$ participants into $\by$. With the covariance parameters fixed, the estimate of $\theta$ is a generalized least-squares estimate and is therefore linear in the outcomes: $\widehat\theta=\bh^{\mathsf T}\by$ for a vector $\bh$ of signed observation weights that depends only on the design and the fitted covariance. This linearity is what lets us pass per-scan attributions through the mixed model. The fixed-effects design $\bM\in\mathbb R^{n\times q}$ has full column rank, where $q$ counts coefficients, and $\mathbf Z$ is the random-effects design. Let $\mathbf c\in\mathbb R^q$ select $\theta$, the group-by-time coefficient. For a positive-definite fitted longitudinal covariance, the precision $\bW$, coefficient map $\bL$ and signed observation weights $\bh$ are
\begin{equation}
 \begin{split}
 \bW&=(\mathbf Z\widehat{\mathbf G}_{\rm all}\mathbf Z^{\mathsf T}
                   +\widehat\sigma^2\mathbf I)^{-1},\\
 \bL&=(\bM^{\mathsf T}\bW\bM)^{-1}\bM^{\mathsf T}\bW,
 \qquad\bh=\bL^{\mathsf T}\mathbf c,
 \end{split}
 \label{eq:weights}
\end{equation}
where $\widehat{\mathbf G}_{\rm all}$ has one $\widehat{\mathbf G}$ block per participant on its block diagonal. The weights $\bh$ incorporate adjustment and visit dependence. Stack IG in $\bA\in\mathbb R^{n\times p}$, with one region or direction per column. Stack corrected references into $\by_0$ and write $\by-\by_0=\bA\mathbf1_p+\mathbf r$, where $\mathbf1_p$ is all ones and $\mathbf r$ records integration error.

For the contributions to sum to $\widehat\theta$, the reference terms $\by_0$ must receive zero total weight, and they do. Since the age at visit $j$ is $a_{ij}=a_{i0}+t_{ij}$, with $a_{i0}$ the baseline age in $\mathbf u_i$, and $f(\bx_0)$ is constant, $\by_0$ lies in the span of the intercept, baseline-age and time columns of $\bM$. Since the target selects none of these coefficients, $\bh^{\mathsf T}\bM=\mathbf c^{\mathsf T}$ implies $\bh^{\mathsf T}\by_0=0$. Applying each scan's weight to all its components gives
\begin{equation}
 \phi_k=\sum_{s=1}^n h_s A_{sk},\qquad
 \widehat\theta=\bh^{\mathsf T}\by
 =\sum_{k=1}^p\phi_k+\bh^{\mathsf T}\mathbf r.
 \label{eq:components}
\end{equation}
Here $s$ indexes scans. Both decompositions recover the slope difference up to integration error. In particular, combining Eqs.~\eqref{eq:spectral} and~\eqref{eq:components} gives the contribution of latent direction $k$:
\begin{equation}
 \phi_k^{\rm eig}=\sum_{s=1}^n h_s
 (\bv_k^{\mathsf T}\bd_s)(\bv_k^{\mathsf T}\bg_s).
 \label{eq:latent_contrast}
\end{equation}
The two projections measure each scan's displacement along that direction and the network's average sensitivity to it; $h_s$ weights their product according to the adjusted longitudinal analysis. Positive $\phi_k$ means greater adjusted annual change for component $k$ in the positive group. For these same observation weights, subtracting chronological age and the linear age correction changes only nuisance coefficients, leaving the group-by-time coefficient unchanged. The allocations still depend on the common reference.

Contributions of different components are correlated because they are computed from the same participants. To estimate their joint uncertainty and account for repeated scans, define residuals $\mathbf E=\bA-\bM\bL\bA$ and participant vectors $\boldsymbol\xi_i=\mathbf E_i^{\mathsf T}\bh_i$, with subscript $i$ selecting that participant's rows. The cluster-robust sandwich covariance with the CR1 small-sample correction~\cite{Cameron2015} is
\begin{equation}
 \widehat{\boldsymbol\Sigma}_{\phi}
 =\frac{N}{N-1}\frac{n-1}{n-q}
 \sum_{i=1}^N\boldsymbol\xi_i\boldsymbol\xi_i^{\mathsf T}.
 \label{eq:uncertainty}
\end{equation}
Intervals for sums retain cross-component covariance. Intervals use $t$ quantiles with $N-1$ degrees of freedom and treat the observation weights and the prediction pipeline (VNN, reference and age correction) as fixed.

\section{Results}
\subsection{VNN pipeline and longitudinal effect to be explained}
\label{sec:data}
Data used in the preparation of this article were obtained from the Alzheimer’s Disease Neuroimaging
Initiative (ADNI) database (adni.loni.usc.edu). The ADNI was launched in 2003 as a public-private
partnership, led by Principal Investigator Michael W. Weiner, MD. The primary goal of ADNI has been to
test whether serial magnetic resonance imaging (MRI), positron emission tomography (PET), other
biological markers, and clinical and neuropsychological assessment can be combined to measure the
progression of mild cognitive impairment (MCI) and early Alzheimer’s disease (AD).

 CN scans from ADNI~\cite{Jack2008} trained the predictor; MCI participants supplied the longitudinal sample. ADNIMERGE2~\cite{ADNIMERGE2} data were downloaded on February 5, 2026, and the UCSFFSX7 cortical-thickness table on January 30, 2026. Inputs were complete FreeSurfer thickness in 68 Desikan--Killiany regions from 3-T magnetic resonance imaging (MRI)~\cite{fischl2012freesurfer,desikan2006automated}. We included scans that passed manual quality control (QC) or had no recorded QC result; other QC categories were excluded. Before any train/evaluation split, neuroCombat~\cite{Fortin2018,Johnson2007} adjusted scanner differences across 6,535 scans from 2,125 participants and 25 manufacturer--model batches. It preserved linear age/sex effects and treated scans as independent.

One baseline scan from each of 934 CN participants gave 747 development and 187 evaluation participants (seed 42). Optuna~\cite{Akiba2019} ran 200 trials, minimizing mean absolute error (MAE) over ten development folds split by participant; standardization and covariance used each fit's training data. VNN layers 1 and 2 had 198 and 134 output channels and 4 and 11 filter taps, respectively; leaky-ReLU slope was 0.125658, without intermediate pooling. Adam~\cite{kingma2015adam} minimized mean squared error (MSE) with learning rate 0.00589194, batch size 64 and float64 arithmetic. Each fold stopped after 20 epochs without validation MSE improvement or 1,000 epochs, restoring the best weights. The median best epoch, 202, set subsequent training duration. Evaluation MAE was 4.81 years before correction. ComBat used evaluation measurements and ages, precluding a fully held-out pipeline evaluation. MAE alone does not establish disease sensitivity~\cite{Jirsaraie2023,Schulz2025}.

Ten-fold out-of-fold predictions on all 934 CN participants supplied the age correction in Eq.~\eqref{eq:gap}. We then trained the final VNN on all 934. The longitudinal sample contained 464 baseline-MCI participants, 2,278 scans and at least two distinct MRI times per participant. We used the AMYLOID\_STATUS labels from the UC Berkeley amyloid PET table (UCBERKELEY\_AMY\_6MM) included in ADNIMERGE2, derived from florbetapir or florbetaben PET by thresholding cortical uptake normalized to the whole cerebellum~\cite{adniPetDocumentation,Royse2021}. PET records passed QC or were not assessed. One-to-one PET--MRI matches within 60 days of baseline gave 253 amyloid-positive and 211 negative participants. We required complete covariates and used all available follow-up scans.

In positive and negative groups, respectively, mean baseline age was 73.6 (SD 6.8) and 70.3 (8.3) years. Median follow-up from baseline to last included MRI was 2.09 and 3.00 years, with interquartile ranges (IQRs) 1.17--4.05 and 1.98--6.80; the overall maximum was 14.92 years. Median scan counts including baseline were 5 (IQR 3--5) and 5 (4--6). Time was expressed in five-year units for numerical conditioning of the mixed-model fit; slope estimates were converted back to annual values. The fitted mixed model estimated an adjusted annual $\Delta$-Age increase 0.375 years/year greater in amyloid-positive than amyloid-negative participants (95\% CR1 CI: 0.153 to 0.597). \textbf{This longitudinal pattern is the target of the attribution analysis below.}

\subsection{Latent and regional contributions to the longitudinal effect}
\label{sec:contrib}
We computed IG by a 512-point midpoint rule from the final VNN's CN training mean, then applied the observation weights from the fitted $\Delta$-Age model. We report each contribution as a percentage of its signed 0.375 years/year slope difference, which we call the net contrast. Figure~\ref{fig:results}a adds signed contributions in order of decreasing absolute value within each coordinate system.

\begin{figure}[t]
 \centering
 \includegraphics[width=0.7\columnwidth]{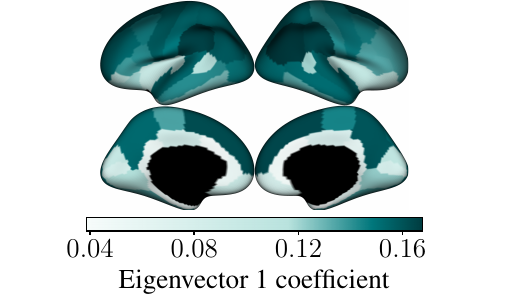}
 \caption{Regional coefficients of the leading training-covariance eigenvector (EV1), in standardized cortical-thickness coordinates. The vector has unit norm and its sign is chosen to make its coefficients positive; reversing that sign leaves IG unchanged. Colors describe the direction itself, not regional IG contributions.}
 \label{fig:ev1}
\end{figure}

In the latent representation, the leading eigenvector (EV1) contributed 0.312 years/year (95\% CR1 CI: 0.173 to 0.451), or 83.2\% of the net contrast. Figure~\ref{fig:ev1} shows the regional coefficients defining this direction. EV1 explains 32.5\% of training variance, a different quantity from its 83.2\% share of the net contrast. EV2--68 jointly contributed 0.0629 years/year (95\% CR1 CI: $-0.0474$ to 0.173), or 16.8\%; the sum of their absolute contributions, $\sum_{k=2}^{68}|\phi_k|=0.219$ years/year, shows cancellation between positive and negative contributions. Regional contributions accumulate more gradually. The largest was left inferior parietal, 0.0290 years/year (95\% CR1 CI: 0.0137 to 0.0443), or 7.7\%; right middle temporal contributed 0.0284 years/year (95\% CR1 CI: 0.0140 to 0.0427), or 7.6\%. The regional decomposition distributes the contrast across individual inputs, whereas the latent decomposition concentrates it along one covariance direction. These are different allocations of the same fitted effect; an eigenvector contribution is not a sum of selected regional IG entries.

Reusing the full $\Delta$-Age model's observation weights makes the contributions reconstruct its fitted group difference, up to integration error. Two natural alternatives do not. The first ignores the mixed model: for each participant we fitted a least-squares slope of each component's IG against time, averaged the slopes within each amyloid group, and took the positive-minus-negative difference. This gives a total of 0.267 years/year (EV1: 0.171), because it drops the covariate adjustment and weights every participant equally regardless of visit count and follow-up. The second fits the adjusted mixed model of Eq.~\eqref{eq:lmm} separately, once with EV1's IG as the outcome and once with the EV2--68 sum, and reads off each $\theta$. The two estimates sum to 0.356 (EV1: 0.294), not 0.375, because each fit estimates its own longitudinal covariance and therefore its own observation weights.

\subsection{Checking IG against the VNN's behavior}
\label{sec:intervention}
To test whether the VNN uses EV1 in the way the attribution suggests, we modified each scan's displacement to keep only its EV1 component or to remove it. With $\mathbf P_1=\bv_1\bv_1^{\mathsf T}$, we evaluated every scan at $\bx_0+\mathbf P_1\bd$ (EV1 only) and $\bx_0+(\mathbf I-\mathbf P_1)\bd$ (EV1 removed), keeping the VNN, reference and observation weights fixed, and recomputed the adjusted group difference from the modified predictions. Removing EV1 sets its displacement from the reference to zero while preserving every other covariance coordinate of the input. We compare the keep-EV1-only contrast with EV1's IG contribution and the removal contrast with the combined contribution of EV2--68. This checks how closely the attribution split matches the fitted VNN's response to retaining or removing that direction. Keeping only EV1 gave 0.296 years/year (95\% CR1 CI: 0.159 to 0.432) and removing it gave 0.0622 (95\% CR1 CI: $-0.0507$ to 0.175) (Fig.~\ref{fig:results}b). The keep-EV1-only contrast minus the EV1 IG contribution was $-0.0164$ (paired 95\% CR1 CI: $-0.0458$ to 0.0129), and the removal contrast minus the EV2--68 contribution was $-0.0007$ (paired 95\% CR1 CI: $-0.0140$ to 0.0126); paired intervals account for the two quantities' shared participants. Refitting the mixed model on the modified predictions, instead of reusing the weights, gave 0.291 for retention and 0.0619 for removal. With the original observation weights, the intervention and IG point estimates differ by at most 0.017 years/year, about 5\% of EV1's contribution. Retaining EV1 preserves most of the contrast, whereas removing it leaves a much smaller contrast. These responses support EV1's dominant contribution to this fitted longitudinal effect. Agreement is not guaranteed by IG completeness, which fixes only the sum of the attributions: changing one covariance component can also change the network's sensitivity to the others.

\subsection{Numerical and fitted-model sensitivity}
In Eq.~\eqref{eq:components}, the reconstruction error was within $2.20\times10^{-5}$ years/year in both coordinate systems; doubling integration points changed EV1 by $4.17\times10^{-6}$. To check EV1's directional stability, 2,048 CN participant bootstraps re-estimated standardization and covariance without retraining or recomputing IG. The absolute dot product between original and resampled EV1 had median 0.9992 and central 95\% range 0.9984--0.9995 (1 is identical up to sign).

For longitudinal uncertainty, we resampled participants with all visits 2,048 times within amyloid groups, refitting the mixed model and recomputing observation weights and contributions; all fits converged. Preprocessing, predictions and IG stayed fixed. EV1 exceeded the signed sum of EV2--68 by 0.249 years/year (95\% percentile bootstrap CI: 0.131 to 0.378). Ten additional VNN fits varied only the initialization and minibatch order, holding preprocessing, reference and age correction fixed. EV1 remained the largest absolute contributor, ranging from 78.8\% to 83.5\% of the net contrast across fits. For preprocessing sensitivity, a separately trained VNN without ComBat, adding scanner-batch main effects to the mixed model, gave total 0.449 (95\% CR1 CI: 0.246 to 0.652) and EV1 contribution 0.393 (95\% CR1 CI: 0.247 to 0.538), or 87.5\%.

\section{Discussion}
The latent decomposition of longitudinal effects in brain age gap have revealed that a single covariance direction carried most of the fitted VNN's adjusted difference in brain age gap progression between amyloid-positive and amyloid-negative individuals with MCI, whereas regional IG spread the same effect across many cortical inputs. Retention and removal interventions have confirmed that the VNN relied on this direction as the attribution indicated. The 83.2$\%$ share of the leading covariance eigenvector in explaining the longitudinal effect suggests that the pre-trained VNN's sensitivity along this direction was instrumental to the longitudinal findings. The analysis was limited to one longitudinal contrast in a single cohort. Future work will apply the framework to other cohorts, clinical contrasts and trajectory models, and will examine whether the dominant latent directions generalize across populations.

\section{Compliance with Ethical Standards}

ADNI obtained institutional ethics approval and written informed consent from
participants or legally authorized representatives. Our secondary analysis
used de-identified data under the ADNI Data Use Agreement, without new
recruitment or data collection. %The authors declare no conflicts of interest.

\section{Acknowledgments}
$^{\ast}$ Data used in preparation of this article were obtained from the Alzheimer’s Disease Neuroimaging Initiative
(ADNI) database (adni.loni.usc.edu). As such, the investigators within the ADNI contributed to the design
and implementation of ADNI and/or provided data but did not participate in analysis or writing of this report.

Data collection and sharing for the Alzheimer's Disease Neuroimaging Initiative (ADNI) is funded by the National Institute on Aging (National Institutes of Health Grant U19AG024904). The grantee organization is the Northern California Institute for Research and Education. In the past, ADNI has also received funding from the National Institute of Biomedical Imaging and Bioengineering, the Canadian Institutes of Health Research, and private sector contributions through the Foundation for the National Institutes of Health (FNIH) including generous contributions from the following: AbbVie, Alzheimer’s Association; Alzheimer’s Drug Discovery Foundation; Araclon Biotech; BioClinica, Inc.; Biogen; BristolMyers Squibb Company; CereSpir, Inc.; Cogstate; Eisai Inc.; Elan Pharmaceuticals, Inc.; Eli Lilly and Company; EuroImmun; F. Hoffmann-La Roche Ltd and its affiliated company Genentech, Inc.; Fujirebio; GE Healthcare; IXICO Ltd.; Janssen Alzheimer Immunotherapy Research \& Development, LLC.; Johnson \& Johnson Pharmaceutical Research \& Development LLC.; Lumosity; Lundbeck; Merck \& Co., Inc.; Meso Scale Diagnostics, LLC.; NeuroRx Research; Neurotrack Technologies; Novartis Pharmaceuticals Corporation; Pfizer Inc.; Piramal Imaging; Servier; Takeda Pharmaceutical Company; and Transition Therapeutics.

\bibliographystyle{IEEEbib}
\bibliography{refs}
\end{document}